\documentstyle[epsfig]{aipproc}
\begin{document}
\title{Synchrotron Radiation as the Source of GRB Spectra, Part I: Theory.}

\author{Nicole M. Lloyd \& Vah\'e Petrosian}
\address{Stanford University,
Stanford, California 94305}

\maketitle

\begin{abstract}
We investigate synchrotron emission models as the source
of gamma ray burst spectra.  We show that allowing for synchrotron self
absorption and a ``smooth cutoff'' to the electron energy distribution
produces a wide range of low energy spectral behavior.  We show that
there exists a correlation between the value of the peak
of the $\nu F_{\nu}$ spectrum, $E_{p}$, and the
low energy spectral index $\alpha$ as determined by spectral
fits over a finite bandwidth.  Finally, we discuss the implications of
synchrotron emission from internal shocks for GRB spectral evolution.
\end{abstract}

\section*{Introduction}
It has been suggested (e.g. \cite{katz94}) that {synchrotron emission} is
a likely source of radiation from GRBs, and later shown \cite{tav96}
that an optically thin synchrotron spectrum is a good fit to some bursts.
However, some features seen in the low
energy portion of GRB spectra can not be explained
by the simple synchrotron model (SSM) - optically thin synchrotron
emission from a power law distribution of relativistic electrons with
a minimum energy cutoff.  This model predicts that the asymptotic
value of the low energy
photon index, $\alpha$,  should be a constant value of $-2/3$. 
 However, the data show an $\alpha$ distribution
with a mean of about $-1.1$ and a
standard deviation of about $1$.
  Furthermore, there are
a significant fraction of bursts with $\alpha > -2/3$ - above the so-called
{``line of death''} \cite{pree98}. 
 In addition, spectral evolution of $\alpha$ and the peak
of $\nu F_{\nu}$, $E_{p}$, are inconsistent with an instantaneous
optically thin synchrotron spectrum in an external shock model \cite{crid97}. 
{Consequentially, other models - usually involving inverse compton scattering
\cite{brain94}, \cite{lian96} -
 were invoked to explain these ``anomolous''
spectral behaviors.}

   In this paper, we discuss how 
GRB spectra can be accomodated by    
{synchrotron emission}, including those
spectra not explained by the SSM.  
We discuss the various spectral shapes from a general form
for synchrotron emission, allowing for the possibility for self-absorption 
and a smooth cutoff to the electron energy distribution, and  
show that these models
fit GRB spectra well.
We show there is a correlation between $\alpha$ and
$E_{p}$ as determined by fits using 
the Band \cite{band93} (and similiar) spectral forms.  
Finally, we briefly discuss the variety of spectral evolution behaviors
seen in GRBs in the context of synchrotron emission.
In Part II (Lloyd et al., these proceedings, \cite{PartII}),
 we compare our theoretical predictions 
with the data and show how synchrotron emission can explain
the spectral behavior of GRBs.
 
\section*{Synchrotron Emission}
  The general form for an instantaneous synchrotron spectrum
for a power law distribution in the electron energy with a {sharp
cutoff}, $N(E) = N_{o}E^{-p}$, $E>E_{min}$, is given by \cite{pach}
\begin{equation}
F_{\nu}=A \nu^{5/2}
[\frac{I_{1}}{I_{2}}] \times  
[1.0 - {\rm exp}[-Q \nu^{-(p+4)/2}
I_{2}]]
\end{equation}
\begin{equation}
I_{1} = \int_{0}^{\frac{\nu}{\nu_{min}}}dx\ x^{(p-1)/2}
\int_{x}^{\infty} K_{5/3}(z) dz, \ I_{2} = \int_{0}^{\frac{\nu}{\nu_{min}}}
dx\ 
x^{p/2}
\int_{x}^{\infty} K_{5/3}(z) dz
\end{equation}
$A$ is the normalization and contains factors involving
the perpendicular
component of the magnetic field, $B_{\perp}$, bulk
Lorentz factor, $\Gamma$, and number of electrons,
$N_{o}$. The frequency
$\nu_{min} = (\Gamma E_{min}^{2} B_{\perp} 3 e)/
(m^{3} 4 \pi c^{2})$. 
The parameter $Q$ represents the optical
depth of the medium (for example, if $\nu \gg \nu_{min}$, the
photon spectrum will be absorbed at the frequency $\nu_{abs} \sim
Q^{2/(p+4)}$).  The high energy asymptotic behavior is the usual $F_{\nu}
 \propto \nu^{-(p-1)/2}$.
The low energy asymptotic forms of the function depend on the relative
values of $\nu_{min}$ and $\nu_{abs}$: 
 $F_{\nu} \sim \nu^{5/2}$ for $\nu_{min} < \nu \ll \nu_{abs}$,
 $F_{\nu} \sim \nu^{2}$ for $\nu \ll {\rm min}[\nu_{abs},\nu_{min}]$,
 $F_{\nu} \sim \nu^{1/3}$, for $\nu_{abs} < \nu < \nu_{min}$.

Note that we do not address the case of cooling electrons,
which will have the effect of increasing the electron power law distribution index
$p$ by 1, $p \rightarrow p+1$ at some characteristic cooling energy.  

\section*{The Electron Distribution}
 In most models of synchrotron emission, the
 electron distribution is modeled by a power law with
 a sharp cutoff at some minimum energy (as done in the previous
 section).  This is not a realistic (and may even be an
 unstable) distribution.
We characterize the electron distribution
by the following equation:
\begin{equation}
N(E) = N_{o}\frac{(E/E_{*})^{q}}{1+(E/E_{*})^{p+q}}
\end{equation}
where $E_{*}$ is some critical energy that characterizes
where the electron distribution changes. 
 For $E \gg E_{*}$, $N(E) \propto E^{-p}$, 
while for $E \ll E_{*}$, $N(E) \propto E^{q}$.  Hence,
 q characterizes the ``smoothness'' of the
cutoff.  This has a significant impact on the low energy
portion of the synchrotron spectrum.
{An optically thin synchrotron spectrum takes the form:
\begin{equation}
F_{\nu} = A(\nu/\nu_{*})^{(q+1)/2}\int_{0}^{\infty}dx
\frac{x^{-(q+p)/2}}{1+((\nu/\nu_{*})^{(q+p)/2}x^{(p+q)/2})}
\int_{x}^{\infty} K_{5/3}(z)dz
\end{equation}
where $\nu_{*} = c_{1}B_{\perp}E_{*}^{2}$.}
Depending on the smoothness of the cutoff,
the  spectrum of the emitted photons can change significantly.
{The peak of the spectrum is shifted to lower energies
as the cutoff becomes smoother (q smaller),} and
{the width of the spectrum increases, which implies that it takes longer
for the spectrum to reach its low energy asymptotic value.}
\begin{figure}
\centerline{\epsfig{file=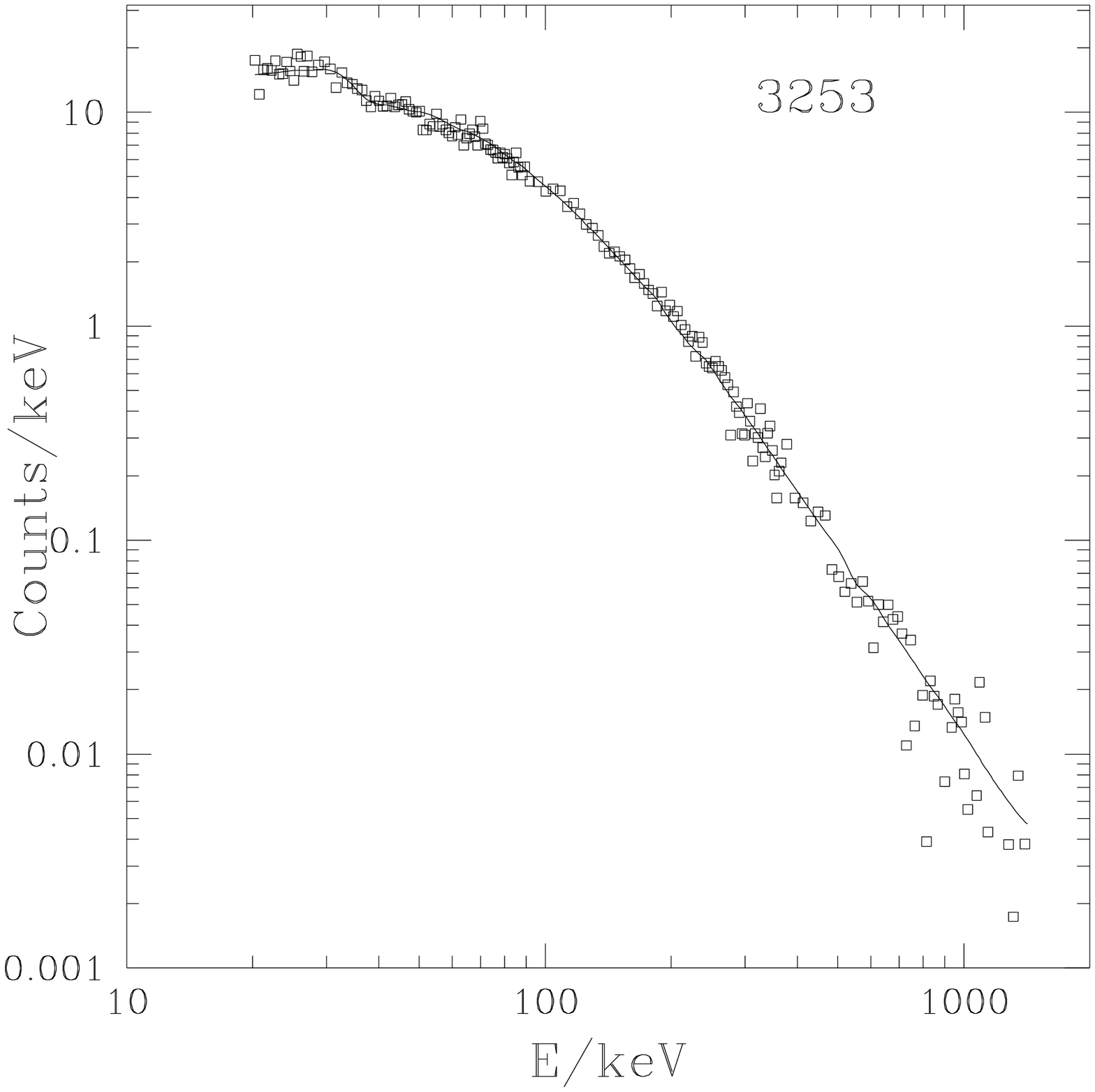,width=.55\textwidth,height=0.45\textwidth}
\epsfig{file=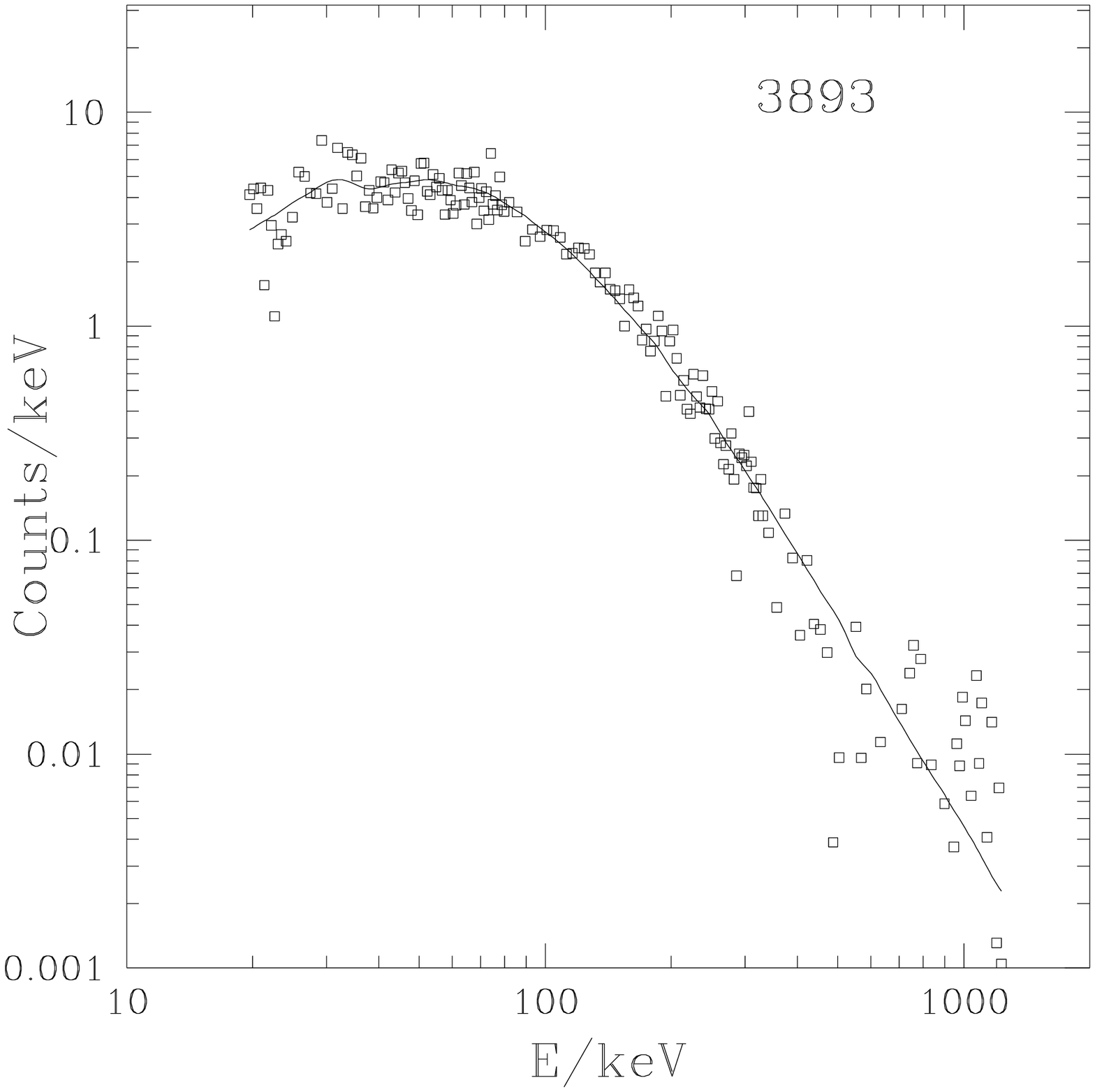,width=.55\textwidth,height=0.45\textwidth}} 
\caption{Spectra of two GRBs; the solid line shows the synchrotron model
fit. An optically thin model is the best fit to 3253 (left), while
and optically thick model best describes 3893 (right). }
\end{figure}

\section*{Spectral Fits}
 Synchrotron emission models
fit the GRB data well.
    We fit 11 bursts with 256 channel energy resolution
to the synchrotron spectral
forms described above. 
   Five of
these bursts have a low energy photon index (as determined
by fitting a Band spectrum)
 above the ``line of death''
for optically thin synchrotron emission; that is, $\alpha>-2/3$. 
 In all of these cases, we found that
{\em including an absorption parameter will accomodate the hardness
of the low energy index}, and provided the best fit.
   { {Figure 1}} shows the spectra
for 2 GRBs in our sample (burst triggers 3893 and 3253).  A self-absorbed
spectrum in which the absorption frequency
just enters the BATSE window best fits 3893, while an optically thin spectrum is the best
fit to 3253.  For a more complete discussion of the spectral fits, see
\cite{inprep}.

\section*{A Range of Spectra}
  Figure 2a shows the many types of low energy spectral behavior one
  can obtain from the above synchrotron models,
normalized to the peak of $F_{\nu}$ (at $500$ keV).
The vertical lines mark the approximate width of the BATSE spectral window.

Now, the $\alpha$ distribution will depend largely on how quickly the
spectrum reaches its low energy asymptote or how well spectral fits
can determine the asymptote.  As 
$E_{p}$ moves to lower and lower energies, we get less and less of
the low energy portion of the spectrum; in this case, our spectral
fits probably will not be able to determine the asymptote and will
measure a lower (softer) value of $\alpha$.  [Preece et al. \cite{pree98}
pointed out this effect and attempt to minimize it by
defining an effective $\alpha$, which is the slope of the spectrum
at 25keV (the edge of the BATSE window).  However, a correlation
between $\alpha_{eff}$ and $E_{p}$ will still exist if the
asymptote is not reached well before 25keV.]  This difficulty
becomes more severe the smoother the cutoff to the electron distribution,
because the spectrum takes longer to reach its asymptote.
To test this, we
produce sets of data from optically
thin synchrotron models with different parameters
($\nu_{min}$, $q$, etc.), all of which have a low
energy asymptote of $-2/3$.  We 
fit a Band spectrum to this data (to be conservative,
we extended the range of BATSE's sensitivity to $10$ keV).
Figure 2b 
shows the value of the asymptote as determined by
the Band spectrum, as a function of $E_{p}$, for
different degrees of the smoothness of the electron energy distribution 
cutoff.
Not surprisingly, {  there is a strong correlation
between the value of $E_{p}$ and the value of
the ``asymptote'', $\alpha$, as determined by a Band fit to
the data}.
 {We can use this relationship and knowledge
of the $E_{p}$ distribution to determine the resultant distribution
for $\alpha$.
This is discussed in Part II \cite{PartII}.}
\begin{figure}
\centerline{\epsfig{file=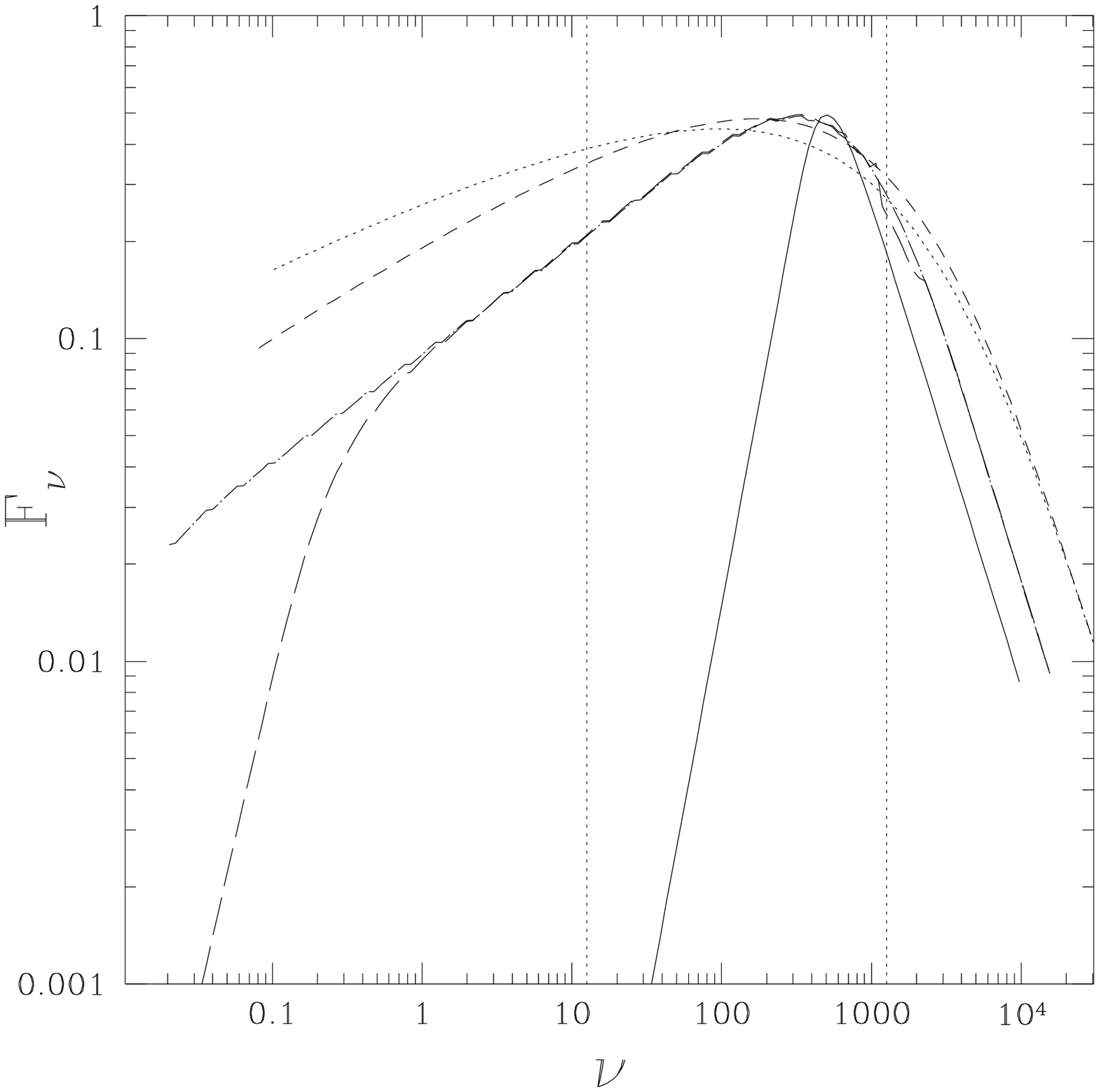,width=.55\textwidth,height=0.45\textwidth}
\epsfig{file=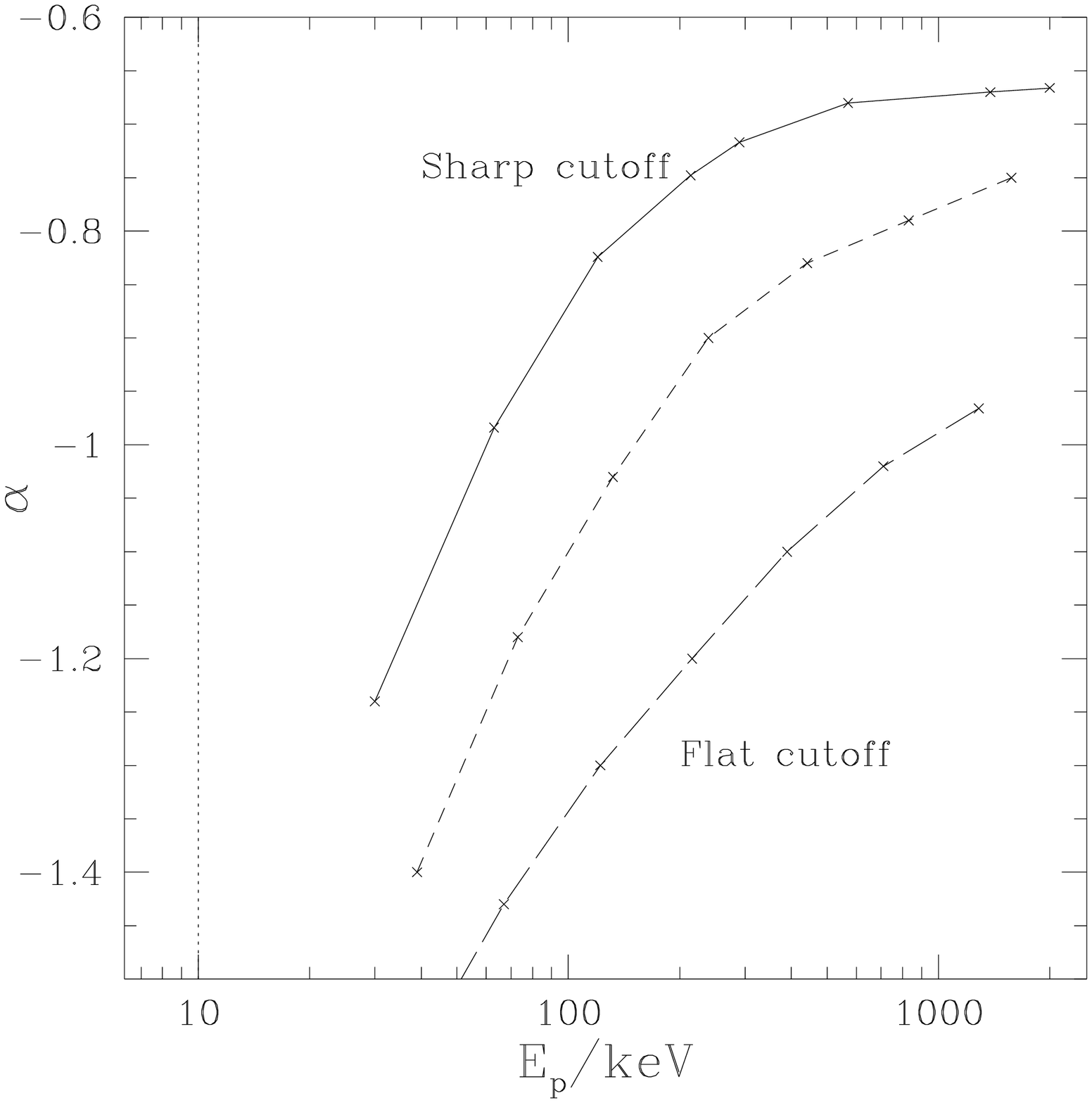,width=.55\textwidth,height=0.45\textwidth}} 
\caption{{\bf Left Panel}: Synchrotron spectra for different values of the
optical depth and smoothness of the electron cutoff. Optically
thin spectra are shown by the the dot-dashed
line, the dotted line, and the short-dashed line 
for a sharp ($q=\infty$, SSM), an intermediate ($q=2$)
and flat ($q=0$) cutoff to the $e^{-}$ distribution, respectively.
The solid and long dashed lines show the self-absorption
cutoff when $\nu_{abs}>\nu_{min}$ and $\nu_{abs}<\nu_{min}$,
respectively.  The vertical
lines mark the BATSE window.  {\bf Right Panel}: The
correlation between $\alpha$ and $E_{p}$ for a spectrum with a sharp (solid
line), intermediate (short dashed), and flat (long dashed) cutoff to the
$e^{-}$ distribution.}
\end{figure}

\section*{Spectral Evolution}
The behavior of the spectral characteristics with time throughout
 the GRB can give us
information
about the environment of the emission region
and conceivably constrain the emission mechanism.  Given the
apparent correlation
between $\alpha$ and $E_{p}$ induced by the fitting
procedure, we expect
evolution of $\alpha$ (obtained from such fits) to mimic the behavior of
 $E_{p}$ in time during a pulse or spike.  Note, however, if
 each pulse in the
time profile is a separate emission episode (as in an internal
shock scenario), parameters such as $q$ and
the optical depth can vary from
shock to shock; this can create a change in $\alpha$ from
pulse to pulse, independent of
$E_{p}$. 

\section*{Conclusions}

Synchrotron emission can produce a
variety of GRB spectral shapes, particularly when one allows for 
a smooth cutoff to the electron distribution and includes effects of
self-absorption.
{In addition, we expect a relationship between $\alpha$ and $E_{p}$, as
 a consequence of the fitting procedure; this will have implications for the  
 observed distributions and temporal evolution of spectral parameters.}
 We test this model against the data in Part II \cite{PartII}.

\end{document}